\documentclass[pss]{wiley2sp} 
\usepackage{amsmath}
\newcommand{\kk}{{\bf k}}
\newcommand{\qq}{{\bf q}}
\newcommand{\rl}{\rangle\!\langle}
\newcommand{\rc}{\rangle^{\mathrm corr}}

\newcommand{\sk}{s_\kk}
\newcommand{\skp}{s_\kk^{(+)}}
\newcommand{\tk}{t_\kk}
\newcommand{\sq}{s_\qq}
\newcommand{\sqp}{s_\qq^{(+)}}
\newcommand{\tq}{t_\qq}

\begin{document}
\title{Phonon-assisted decoherence and tunneling in quantum dot molecules}
\titlerunning{Phonon-assisted decoherence and tunneling in QDMs}
\author{%
  Anna Grodecka-Grad\textsuperscript{\Ast,\textsf{\bfseries 1,2}} and
  Jens F{\"o}rstner\textsuperscript{\textsf{\bfseries 2}}}
\authorrunning{Anna Grodecka-Grad et al.}
\mail{e-mail
  \textsf{anna.grodecka-grad@nbi.dk}, Phone:
  +45-353-25426, Fax: +45-353-25400}
\institute{%
  \textsuperscript{1}\,QUANTOP, Danish National Research Foundation Center for Quantum Optics, 
  Niels Bohr Institute, University of Copenhagen, DK-2100 Copenhagen {\O}, Denmark\\
  \textsuperscript{2}\,Computational Nanophotonics Group, Theoretical Physics, University of Paderborn, 33098 Paderborn, Germany}

\received{XXXX, revised XXXX, accepted XXXX} 
\published{XXXX} 


\abstract{%
\abstcol{%
We study the influence of the phonon environment on the electron dynamics in a doped quantum dot molecule.
A~non-perturbative quantum kinetic theory based on correlation expansion is used in order to describe
both diagonal and off-diagonal electron-phonon couplings representing real and virtual processes
with relevant acoustic phonons. 
}{%
We show that the relaxation is dominated by phonon-assisted electron tunneling between constituent quantum dots
and occurs on a picosecond time scale. The dependence of the time evolution of the quantum dot occupation probabilities
on the energy mismatch between the quantum dots is studied in detail.}}

\maketitle   

\section{Introduction}	

Quantum dots (QDs) are one of the promising candidates for building a feasible quantum computer.
In particular, it has been proposed to use two coupled quantum dots called quantum dot molecules (QDMs)
for various schemes of quantum computation, where the two ground states of a single confined electron
as well as singlet and triplet states of doubly doped structures can be used as the logical qubit states.
The electrical \cite{petta05} and optical control \cite{bayer01,hakan07} of spin- and charge-based qubits
have already been demonstrated.
However, in such solid state systems, the phonon-assisted relaxation \cite{brandes99,debald02} can strongly affect the coherent control.
The energy difference between the two electron ground states, which is typically of order of a few meV,
can lead to pure dephasing processes \cite{grodecka07,machnikowski04b,grodecka09}
and to electron tunneling \cite{grodecka08,grodecka10,gawarecki10}. 

In this paper, we present the full quantum kinetic description of the phonon-mediated
relaxation in doped quantum dot molecules including non-Markovian effects.
We show that the coupling to the phonon reservoir in quantum dot molecules can lead
to a fast electron tunneling on a picosecond timescale, which strongly affects the coherent electron evolution.
We employ the non-Markovian correlation expansion technique \cite{rossi02,forstner03} including up to three-particle correlations.
Due to space constrains, we present here equations of motions with up to two-particle correlations.
We include diagonal and off-diagonal couplings to the phonon reservoir representing virtual and real phonon-assisted processes, respectively. 
The dependence of the time evolution of the quantum dot occupation probabilities on the energy mismatch between the constituent
quantum dots is studied in detail. 

\section{Model system}

We consider a system consisting of a single quantum dot molecule doped with one electron. The relaxation between
the two energetically lowest states of the electron $|1\rangle$ in the left and $|2\rangle$ in the right quantum dot is considered 
(see Fig.~1 with schematic plot of the energy levels in the quantum dot molecule).
The free Hamiltonian of the electron reads
\begin{equation}
H_{\mathrm{c}} = \epsilon \left( |2\rl 2| - |1\rl 1| \right) + \Gamma \left(|1\rl 2| + |2 \rl 1 | \right),
\end{equation}
where $\Gamma$ is the direct tunneling coupling element between the quantum dots and $\Delta \epsilon = 2\epsilon$ is the energy difference
between the ground states in both quantum dots.

\begin{figure}[h]
\begin{center}
\includegraphics*[width=0.8\linewidth]{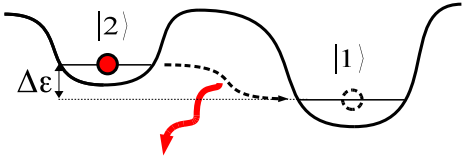}
\caption{A schematic plot of a quantum dot molecule with two electron states
$|1\rangle$ and $|2\rangle$ with the energy difference of $\Delta \epsilon = 2 \epsilon$. The dashed and curled arrows depict electron
tunneling with simultaneous phonon emission.}
\end{center}
\end{figure}

The Hamiltonian describing the free evolution of phonons is
\begin{equation}
H_{\mathrm{ph}} = \sum_{\kk,s} \hbar \omega_{\kk,s} b_{\kk,s}^\dag  b_{\kk,s}
\end{equation}
with $b_{\kk,s}^\dag$ and $b_{\kk,s}$ being phonon creation and annihilation operators, respectively.
$\omega_{\kk,s}$ denotes phonon frequencies with phonon wave vector $\kk$, where different phonon branches
are labeled by $s$ (longitudinal $l$ and two transverse $t1$ and $t2$).

The interaction between the electron and phonon reservoir reads
\begin{equation}
H_{\mathrm{int}} = \sum_{i,j=1,2} \sum_{\kk,s} \left[ g_{ij,s} (\kk) |i\rl j| b_{\kk,s} + 
g_{ij,s}^* (\kk) |j\rl i| b_{\kk,s}^\dag \right].
\end{equation}
The relevant coupling elements for the acoustic phonons coupled via both piezoelectric coupling and deformation potential read:
\begin{equation}
g_{ij,s}^{\mathrm{PE}}(\kk) = -i \sqrt{ \frac{\hbar}{2 \rho c_s k} } 
\frac{d_P e}{\varepsilon_0 \varepsilon_r} M_s(\hat k)
\int d^3r \; \psi_i^*({\bf r}) e^{i\kk \cdot {\bf r}} \psi_j ({\bf r}),
\end{equation}
\begin{equation}
g_{ij,l}^{\mathrm{DP}}(\kk) = \sqrt{ \frac{\hbar k}{2 \rho V c_l} }  D_{e} 
\int d^3r \; \psi_i^*({\bf r}) e^{i\kk \cdot {\bf r}} \psi_j ({\bf r})
\end{equation}
with Gaussian electron wave functions
\begin{equation}
\psi_i({\bf r}) = \frac{1}{\pi^{3/4} l_i^{3/2}} \exp\left({-\frac{{\bf r}^2}{2 l_i^2}}\right).
\end{equation}
We have numerically simulated the self-assembled GaAs quantum dots with the following parameters:
$l_1 = 4$~nm, $l_2 = 4.1$~nm, and $d = 6$~nm, where $l_i$ is the electron wave function size of the $i$-th quantum dot, 
and the distance between the dots is~$d$.
Here, $\rho = 5360$~kg/m$^3$ is the crystal density, $V$ is the normalization volume of the phonon modes,
$c_s$ is the speed of sound (longitudinal $c_l = 5150$~m/s or transverse $c_t = 2800$~m/s, depending on the phonon branch),
$d_P = 0.16$~C/m$^2$ is the piezoelectric constant, $D_e = -8$~eV is the deformation potential constant 
for the electrons and $\varepsilon_r = 13.2$ is the static dielectric constant. 
The function $M_s(\hat k)$ in the piezoelectric coupling element reads:
\begin{eqnarray} 
M_l(\hat k) & = & \frac{3}{2} \sin(\theta) \sin(2\theta) \sin(2\varphi), \\
M_{t1}(\hat k) & = & - \sin(2\theta) \cos(2\varphi),\\
M_{t2}(\hat k) & = & \sin(\theta) [3\cos^2(\varphi)-1] \sin(2\varphi).
\end{eqnarray}

We study the time evolution of the electron confined in a quantum dot molecule
in the presence of the coupling to the phonon reservoir within the density matrix theory.
We employ the second order correlation expansion method \cite{rossi02,forstner03},
where it is assumed that correlations involving an increasing number of particles are of decreasing importance. 
This non-perturbative technique covers the memory effects in the non-Markovian regime.

With the help of Heisenberg equation we derive the equations of motion for the quantities of interest.
The equation for the electron occupation $f = \langle | 2 \rl 2 | \rangle$ reads:
\begin{eqnarray}\nonumber
\dot f & = & \frac{2}{\hbar} \Gamma \; \mathrm{Im} (p)
- \frac{i}{\hbar} \sum_\kk g_{12}(\kk) \left[\sk - \sk^{(+)*} + 2i B_\kk \mathrm{Im} (p) \right] \\ 
&&- \frac{i}{\hbar} \sum_\kk g_{12}^*(\kk) \left[\skp - \sk^{*} + 2i B_\kk^* \mathrm{Im} (p) \right] 
\end{eqnarray}
It couples to the coherence $p = \langle | 1 \rl 2 | \rangle$ as well as to the phonon-assisted coherences
$\sq = \langle |1 \rl 2| b_\qq \rc$ and $\sqp = \langle |1 \rl 2| b_\qq^\dag \rc$ and to phonon amplitudes $B_{\kk} = \langle b_\kk \rangle$.
The factorization scheme has been used above with
\begin{equation}
\langle |1 \rl 2| b_\kk \rangle =  \langle |1\rl 2|\rangle \langle b_{\kk} \rangle +\langle |1 \rl 2| b_\kk \rc,
\end{equation}
where the quantities have been decomposed into all possible lower-order factorizations.
The occupation is directly affected by phonon reservoir in the case of the off-diagonal coupling $\sim g_{12}(\kk) $
and indirectly by the coherence $p$, which evolves in the following way:
\begin{eqnarray}
\dot p & = & - \frac{i}{\hbar}2 \epsilon p - \frac{i}{\hbar} \Gamma (1-2f) \\ \nonumber
&& -\frac{i}{\hbar} \sum_\kk  g_{12}(\kk) \left[ 2\tk + (1-2f)B_\kk \right] \\ \nonumber
&& - \frac{i}{\hbar} \sum_\kk  g_{12}^*(\kk) \left[ 2\tk^* + (1-2f)B_\kk^* \right] \\ \nonumber
&& - \frac{i}{\hbar} \sum_{\kk} \left[ g_{22}(\kk) - g_{11}(\kk) \right] (\sk + p B_\kk) \\ \nonumber
&& - \frac{i}{\hbar} \sum_{\kk} \left[ g_{22}^*(\kk) - g_{11}^*(\kk) \right] \left(\skp + p B_\kk^*\right).
\end{eqnarray}
Here, a direct impact from phonons is present not only via real transitions but also for the pure dephasing 
described by the diagonal coupling elements $g_{11}(\kk)$ and $g_{22}(\kk)$. 
If these coupling elements are equal, the two pure dephasing channels cancel each other out . In this case, the information cannot leak out,
since the phonons is not delivering any \textit{which way} information and no decoherence channel exists.

The evolution of phonons reveals the diagonal and off-diagonal couplings to the density and coherence, respectively,
\begin{eqnarray}
\dot B_\qq & = & -i \omega_\qq B_\qq - \frac{i}{\hbar} g_{22}^*(\qq)
- \frac{i}{\hbar} g_{12}^*(\qq) 2 \mathrm{Re}(p) \\ \nonumber
&& + \frac{i}{\hbar} [ g_{22}^*(\qq) - g_{11}^*(\qq)] (1-f).
\end{eqnarray}

The next step is to derive the equations of motion for all the phonon-assisted correlation quantities.
\begin{eqnarray}
\dot  \sq & = &- \frac{i}{\hbar} 2 \Gamma \tq - \frac{i}{\hbar} \left[ g_{22}^*(\qq) - g_{11}^*(\qq) \right] f p \\ \nonumber
&& - \frac{i}{\hbar}(2 \epsilon + \hbar \omega_\qq) \sq - \frac{i}{\hbar} g_{12}^*(\qq) [1-f - 2 \mathrm{Re}(p)p]  \\ \nonumber
&& - \frac{i}{\hbar} \sum_\kk g_{12}(\kk) \left[ (1-2f) n_{\kk\qq}^{(-)} + 2 \tq B_\kk \right] \\ \nonumber
&& - \frac{i}{\hbar} \sum_\kk g_{12}^*(\kk) \left[ (1-2f) n_{\kk\qq} + 2 \tq B_\kk^* \right] \\ \nonumber &&
- \frac{i}{\hbar} \sum_\kk \left[ g_{22}(\kk) - g_{11}(\kk) \right] \left(
p n_{\kk\qq}^{(-)} + \sq B_\kk \right) \\ \nonumber
&& - \frac{i}{\hbar} \sum_\kk \left[ g_{22}^*(\kk) - g_{11}^*(\kk) \right] \left(
p n_{\kk\qq} + \sq B_\kk^* \right),
\end{eqnarray}
\begin{eqnarray}
\dot \sqp & = &- \frac{i}{\hbar} 2 \Gamma \tq^*- \frac{i}{\hbar} g_{12}(\qq) [ 2 \mathrm{Re}(p) p -f]  \\ \nonumber
&&  - \frac{i}{\hbar}(2 \epsilon-\hbar \omega_\qq ) \sqp  - \frac{i}{\hbar} [g_{22}(\qq) - g_{11}(\qq) ] f p   \\ \nonumber
&& - \frac{i}{\hbar} \sum_\kk g_{12}(\kk) \left[ (1-2f) n_{\qq\kk} + 2 \tq^* B_\kk \right] \\ \nonumber
&& - \frac{i}{\hbar} \sum_\kk g_{12}^*(\kk) \left[ (1-2f) n_{\kk\qq}^{(-)*} + 2 \tq^* B_\kk^*  \right]  \\ \nonumber &&
- \frac{i}{\hbar} \sum_\kk \left[ g_{22}(\kk) - g_{11}(\kk) \right] \left( p n_{\qq\kk} + \sqp B_\kk  \right) \\ \nonumber
&& - \frac{i}{\hbar} \sum_\kk \left[ g_{22}^*(\kk) - g_{11}^*(\kk) \right] \left( p n_{\qq\kk}^{(-)*} + \sqp B_\kk^* \right).
\end{eqnarray}
The dynamics of the phonon-assisted density, $\tq = \langle |1 \rl 1| b_\qq \rc$ is described by
\begin{eqnarray}
\dot \tq & = & -i \omega_\qq \tq - \frac{i}{\hbar} g_{12}^*(\qq) \left[p - 2 \mathrm{Re}(p) (1-f) \right] \\ \nonumber
&& - \frac{i}{\hbar} \Gamma \left[ \sq - \sq^{(+)*} \right] + \frac{i}{\hbar} \left[ g_{22}^*(\qq) - g_{11}^*(\qq) \right] f(1-f) \\ \nonumber
&& - \frac{i}{\hbar} \sum_\kk g_{12}(\kk) \left[ \left(\sq - \sq^{(+)*}\right) B_\kk
+ n_{\kk\qq}^{(-)} 2 i\mathrm{Im}(p)  \right] \\ \nonumber
&& - \frac{i}{\hbar} \sum_\kk g_{12}^*(\kk) \left[ \left(\sq - \sq^{(+)*}\right) B_\kk^* + n_{\kk\qq} 2i \mathrm{Im}(p) \right].
\end{eqnarray}
All these phonon-assisted two-particle correlations couple to the phonon-phonon correlations, 
phonon density $n_{\qq \kk} = \langle b_\qq^\dag b_\kk \rc$ and phonon coherence $n_{\kk \qq}^{(-)} = \langle b_\kk b_\qq \rc $:
\begin{eqnarray}
\dot n_{\qq\kk} & = & -i (\omega_\kk - \omega_\qq)n_{\qq\kk}
+ \frac{i}{\hbar} g_{12}(\qq) \left[\sk + \sk^{(+)*}\right] \\ \nonumber
&& - \frac{i}{\hbar} g_{12}^*(\kk) \left[\sq^* + \sqp \right] + \frac{i}{\hbar} \left[ g_{22}^*(\kk) - g_{11}^*(\kk) \right] \tq^* \\ \nonumber &&
- \frac{i}{\hbar} \left[ g_{22}(\qq) - g_{11}(\qq) \right] \tk,
\end{eqnarray}
\begin{eqnarray}
\dot n_{\qq\kk}^{(-)} & = & -i (\omega_\kk + \omega_\qq)n_{\qq\kk}^{(-)}
- \frac{i}{\hbar} g_{12}^*(\qq) \left[\sk + \sk^{(+)*}\right] \\ \nonumber
&& - \frac{i}{\hbar} g_{12}^*(\kk) \left[\sq + \sq^{(+)*} \right] + \frac{i}{\hbar} \left[ g_{22}^*(\kk) - g_{11}^*(\kk) \right] \tq \\ \nonumber &&
+ \frac{i}{\hbar} \left[ g_{22}^*(\qq) - g_{11}^*(\qq) \right] \tk.
\end{eqnarray}
In addition, the two-particle correlations couple to three particle correlations, e.g.  $\langle |1\rl 2| b_{\qq,s}^\dag b_{\kk,s'} \rc$.
and these couple to up to four particle correlations, etc. Thus, in order to get a closed set of equations,
one needs to truncate the hierarchy by neglecting higher order correlations. 
In the present paper, up to three particle correlations were included.

\section{Phonon-mediated tunneling}

Initially, one electron is injected into the quantum dot with higher energy, $f = \langle |2\rl 2| \rangle =1$.
The time of the injection is assumed to be much shorter than the response of the phonon reservoir,
thus at the initial time we can set all correlations to zero.

\begin{figure}[b]%
\includegraphics*[width=\linewidth]{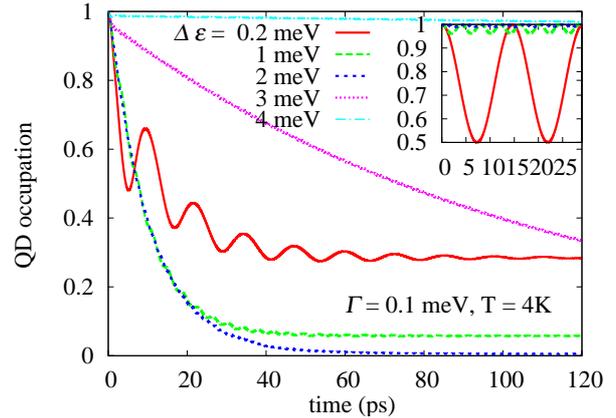}
\caption{\label{fig:e}The occupation probability of a quantum dot with higher energy as a function of time for different
values of the energy difference $\Delta \epsilon$ between two electron ground states for a fixed tunneling coupling $\Gamma = 0.1$~meV
and at fixed temperature $T=4$~K.}
\label{onecolumnfigure}
\end{figure}

The time evolution of the occupation probability of the quantum dot with higher energy 
in the presence of the coupling to the phonon reservoir is shown in Fig.~\ref{fig:e} for different values
of the energy difference $\Delta \epsilon$ between the two electron ground states.
The case of the ideal evolution, where the electron-phonon coupling is absent,
is shown in the inset of Fig.~\ref{fig:e}. The electron tunnels coherently between the two quantum dots
with a period of these oscillations that is determined by the energy difference between the ground states
$\Delta\epsilon = 2 \epsilon$ and the tunneling coupling $\Gamma$ and scales like $\sim\Gamma/(\Delta\epsilon)$.
The maximum occupation also depends on these parameters. 
If the energy difference is twice the value of the tunneling coupling, the electron tunnels between the quantum dots
with maximum occupation of $\frac{1}{2}$ of the quantum dot with lower energy.
For the larger energy differences, the tunneling coupling is too weak to drive the electron into the adjacent quantum dot
and the electron stays in the initial quantum dot.

The dynamics strongly changes if the coupling between the electron and the phonon reservoir is present (see Fig.~\ref{fig:e}),
which results in the electron tunneling to the neighboring quantum dot on a short picosecond time scale.
This phonon-assisted tunneling is only properly modeled if the off-diagonal electron-phonon coupling term in the Hamiltonian is included.
Both coupling via deformation potential and piezoelectric effect are important and contribute to the relaxation.
In the case of the energy difference of $\Delta \epsilon = 0.2$~meV, the thermalization process takes about $50$~ps 
and finishes with a delocalized state of the electron with a probability of $\sim 0.7$ for being in the left quantum dot 
and $\sim 0.3$ in the second dot. For larger energy mismatches of $1$ and $2$~meV, the relaxation takes about $30$ and $40$
picoseconds, respectively, and in the final state, the electron is localized in the quantum dot with lower energy.
In these cases, the probability of phonon emission or/and absorption processes is the highest since
the energies of the relevant acoustic phonons lie in this parameter regime. 
If we increase the energy difference between the dots, the probability of phonon-mediated relaxation 
will decrease and the tunneling is slower and less efficient. 
The presented results were calculated for the low temperature of $T=4$~K.
The phonon-assisted tunneling will get stronger and faster at higher temperatures and will also affect
the final state, since the ratio of probabilities of phonon emission and absorption changes with temperature.

\section{Conclusion}\label{sec4}

A full description of the electron dynamics in the presence of the electron-phonon coupling 
in a quantum dot molecule doped with a single electron has been presented.
The two couplings to the relevant acoustic phonons via deformation potential and piezoelectric coupling
with diagonal and off-diagonal interactions have been taken into account.
It has been shown that the phonon-mediated relaxation is a fast process on a picosecond timescale
strongly modifying the coherent evolution of the electron. It is dominated by off-diagonal electron-phonon coupling.
We analyzed the dependence of the relaxation on the energy difference between the two quantum dots
and indicated the vales of parameters when the phonon-mediated tunneling is most efficient.

\begin{acknowledgement}
The authors acknowledge support from the Emmy Noether Program of the DFG 
(Grant No. FO 637/ 1-1) and the DFG Research Training Group GRK 1464.
\end{acknowledgement}

\end{document}